\documentclass[a4paper,12pt]{iopart}

\expandafter\let\csname equation*\endcsname\relax
\expandafter\let\csname endequation*\endcsname\relax

\usepackage{amsmath}
\usepackage{hyperref}
\usepackage{cite}
\usepackage{graphicx}
\usepackage{amsfonts}
\usepackage{amsthm}
\usepackage{cases}
\usepackage{bm}
\usepackage{subcaption}

\usepackage{fixmath}

\usepackage{enumitem} 

\usepackage[normalem]{ulem}

\usepackage{color}
\definecolor{Blue}{rgb}{0.00, 0.00, 0.80}
\definecolor{Red}{rgb}{0.80, 0.00, 0.00}
\definecolor{Green}{rgb}{0.00, 0.50, 0.00}

\DeclareMathOperator\erfc{erfc}
\DeclareMathOperator\erf{erf}

\hypersetup{
    colorlinks=true,       
    linkcolor=red,          
    citecolor=blue,        
    filecolor=magenta,      
    urlcolor=cyan           
}

\newcommand{\be}{\begin{equation}}
\newcommand{\ee}{\end{equation}}
\newcommand{\bea}{\begin{eqnarray}}
\newcommand{\eea}{\end{eqnarray}}

\newcommand{\beq}{\begin{equation}}
\newcommand{\eeq}{\end{equation}}
\newcommand{\beqn}{\begin{eqnarray}}
\newcommand{\eeqn}{\end{eqnarray}}

\newcommand{\abs}[1]{\ensuremath{\left| #1 \right|}}

\begin{document}

\title{The cost of stochastic resetting}

\author{John C. Sunil$^1$\footnote{Corresponding Author}, Richard A. Blythe$^1$, Martin R. Evans$^1$ and Satya N. Majumdar$^2$}
\address{$^1$ SUPA, School of Physics and Astronomy, University of Edinburgh, Peter Guthrie Tait Road, Edinburgh EH9 3FD, UK}
\address{$^2$ Universit{\'e} Paris-Saclay, CNRS, LPTMS, 91405, Orsay, France}
\eads{\mailto{j.chakanal-sunil@sms.ed.ac.uk}, \mailto{r.a.blythe@ed.ac.uk}, \mailto{m.evans@ed.ac.uk}, \mailto{satya.majumdar@universite-paris-saclay.fr}}

\begin{abstract}
Resetting a stochastic process has been shown to expedite the completion time of some complex  tasks,  such as finding a target for the first time.  Here we consider the cost of resetting by associating to each reset a cost,  which is a function of the distance travelled during the reset event. We compute the Laplace transform of the joint probability of first passage time $t_f$, number of resets $N$ and total resetting cost $C$, and use this to study the statistics of the total cost and also the time to completion ${\mathcal T} = C + t_f$. We show that in the limit of zero resetting rate, the mean total cost is finite for a linear cost function, vanishes for a sub-linear cost function and diverges for a super-linear cost function. This  result contrasts with the case of no resetting where the cost is always zero. We also find that the resetting rate
which optimizes the mean time to completion may be increased or decreased with respect to the case of no resetting cost according to the choice of cost function.
For the case of an exponentially increasing cost function, we show that the mean total cost diverges at a finite resetting rate. We explain this by showing that the distribution of the cost has a power-law tail with a continuously varying exponent that depends on the resetting rate.

\end{abstract}

\submitto{\jpa}
\maketitle

\section{Introduction}
It is well known that resetting a stochastic process to its initial condition can
drastically alter some key properties of the process \cite{EMS20}.
For example, the time to find a target (the first-passage time of the process) can be rendered finite, rather than infinite, by introducing resetting to a diffusive search process \cite{EM1,EM2}. Conditions have been derived for resetting to expedite a process \cite{Reuveni16,PR17} and the explanation is that resetting cuts off errant trajectories that would otherwise create a long time tail in the distribution of completion times.
Many further aspects of the effect of resetting on stochastic processes have been explored in recent years
(see e.g. 
\cite{MV13,EM14,BS14,KMSS14,BEM17,FE17,EM18,Bres20,Bres21,EMS22,
Bres22,Grange22,DBM23}).

One key feature that needs to be taken into account when evaluating the efficacy of resetting is its cost---in the real world resetting can't be instantaneous and  it must consume some resource.
Thus the cost of resetting  encompasses  a number of  possibilities. For example, a time penalty for resetting may be incurred,  either through a refractory period \cite{RUK14,RRU15,EM19}  after the reset, or a return phase of the process  to its resetting position \cite{PKR20,BS20a,BS20b,Radice22}.
The effects of  time penalties on the stationary state and the mean time to find a target have been studied in these works.
In addition, energetic costs can be considered particularly with regard to experimental realizations of resettings \cite{FPSRR20, BBPMC20, FBPCM21}, and even the thermodynamic cost has been appraised \cite{FGS16,MOK23}.

In this work, we consider a general additive cost of resetting where the contribution of each reset is a function of the distance the particle must travel to its resetting position.
Thus  the cost function $C$ for a trajectory involving $N$ resets is
\begin{equation}
  C = \sum_{i=1}^N c_i\;. \label{Cdef}
\end{equation}
Here  $c_i =c(|x_i-x_0|)$ represents the cost of reset $i$, where $x_i$ is the position just before the $i^{\rm th}$ reset and $x_0$ is the resetting position. We use $C$ to represent the total cost over the entire trajectory and $c$ to represent the cost incurred at each reset.
A constant cost $c_i = \mbox{constant}$ corresponds to the case of a refractory period after a reset \cite{RUK14,RRU15,EM19}  and a linear cost $c_i = \frac{1}{V}|x_i-x_0|$ recovers the time penalty incurred by a return phase in which there is a constant velocity $V$ \cite{PKR20,BS20a,BS20b,Radice22}. 
However, in different physical contexts, the cost could vary arbitrarily with the reset distance. For example, one could consider the energy cost, which would depend on the resetting protocol and how energy is dissipated in the surrounding medium. In the remainder of this work, we will not concern ourselves with the details of such processes, instead our aim is to determine the possible behaviours that may emerge by exploring different functional forms of the cost $c(|x_i-x_0|)$ per reset. We suggest possible relevant contexts for the different functional forms.
 We note in passing that the total resetting cost \eqref{Cdef}  is an additive functional. Other additive functionals have been studied in the context of Brownian motion \cite{Majumdar07,MMV23} and Brownian motion with resetting \cite{dHMMT19,SP22}.
   
We will show that different forms of  $c(|x_i-x_0|)$ imply different regimes for the mean total cost for a diffusive process in one dimension under Poissonian resetting with a target at the origin \cite{EM1}.
In particular, as the resetting rate $r\to0$, we show that the mean total cost may be zero, finite or divergent according to the functional form of $c(|x_i-x_0|)$. The cases where the mean total cost is non-zero in the limit of zero resetting are counter-intuitive since one would naively expect zero cost. However, any amount of resetting drastically alters the system's properties, so the limit of zero resetting rate is not equivalent to no resetting.
The explanation of a finite cost is that resetting events with probability $\mathcal{O}(r^{1/2})$ may contribute a cost $\mathcal{O}(1/r^{1/2})$ thus generating a finite mean total cost as $r\to 0$.

In addition, we show that for finite resetting rate $r>0$, the mean total cost may be finite or divergent
according to how quickly the cost function increases with the distance of the reset.
A divergent mean total cost originates in  integrating over a tail of rare events where the resetting distance becomes large.

We can also consider the sum of the total cost, $C$, and the first passage time, $t_f$, which we will refer to as the completion time $\mathcal{T}$
\begin{equation}
    \mathcal{T} = C + t_f\;.
    \label{Tdef}
\end{equation}
We require $C$ to be normalized to have the dimension of time. Then the completion time can be interpreted as the time penalties or return times associated with resetting events, added to the first passage time $t_f$ as has been studied in \cite{PKR20,BS20a}.
We will show that minimising the mean completion time
results in:  a reduced optimal resetting rate, as compared to simply minimising $t_f$, for the case of a linear cost; an unchanged optimal resetting rate for a quadratic cost, and an increased optimal resetting rate for  a super-quadratic cost. In the latter case, the increased optimal resetting rate helps eliminate trajectories with large cost events.
 
The paper is organized as follows. In \sref{sec:renew} we first write down the general renewal equation for the joint probability distribution function of number of resets, time of absorption and cost. We show how various moments of any given cost can be calculated. Then, considering the process with ordinary diffusion in one dimension with a target at the origin, we explore in sections \ref{sec:lin}, \ref{sec:quad}, \ref{sec:power}, \ref{sec-exp} the cases of linear cost, quadratic cost, a generalized power cost and exponential cost, respectively. In particular, we concentrate on various limiting behaviours of each of these costs and the transition from a finite to an infinite cost in various limiting regimes.

\section{Renewal Equation for Cost of Resetting}
\label{sec:renew}

The central quantity we consider is
 the joint probability  distribution function $F(N,t_f,C|x_0)$  of the number of resets $N$, time of absorption $t_f$ and total cost $C$, given that the particle initially starts and is stochastically reset to $x_0$. 
 We will use a renewal approach to compute the  Laplace transform (or moment-generating function) of these joint probabilities
 and hence obtain the statistical properties of the cost. We note that Laplace transforms of related quantities have previously been obtained in
 \cite{BS20a,PKR20}.

We first consider an arbitrary process in one dimension with Poissonian resetting \cite{EMS20} and construct a renewal equation for the  joint probability density by considering the time of the first reset. Related first-renewal equations have been previously used \cite{RUK14,RRU15,PKE16}, and in other contexts, last-renewal equations and  unified renewal approaches have  been employed \cite{EM1,MSS15,PR17,CS18,MPCM19,SGS22,MPCM22}. We find
  \begin{eqnarray}
    F(N,t_f,C|x_0) &=& \int_0^{t_f} {\rm d} t_1 \,r {\rm e}^{-rt_1} \int_0^\infty {\rm d} x_1\,
                       G_0(x_1, t_1|x_0) \label{renew} \\
    &&\times F(N-1,t_f - t_1,C -c_1|x_0) \Theta(C-c_1)  \;,\nonumber
  \end{eqnarray}
  \begin{figure}[h]
    \centering
    \includegraphics[width = 0.8\textwidth]{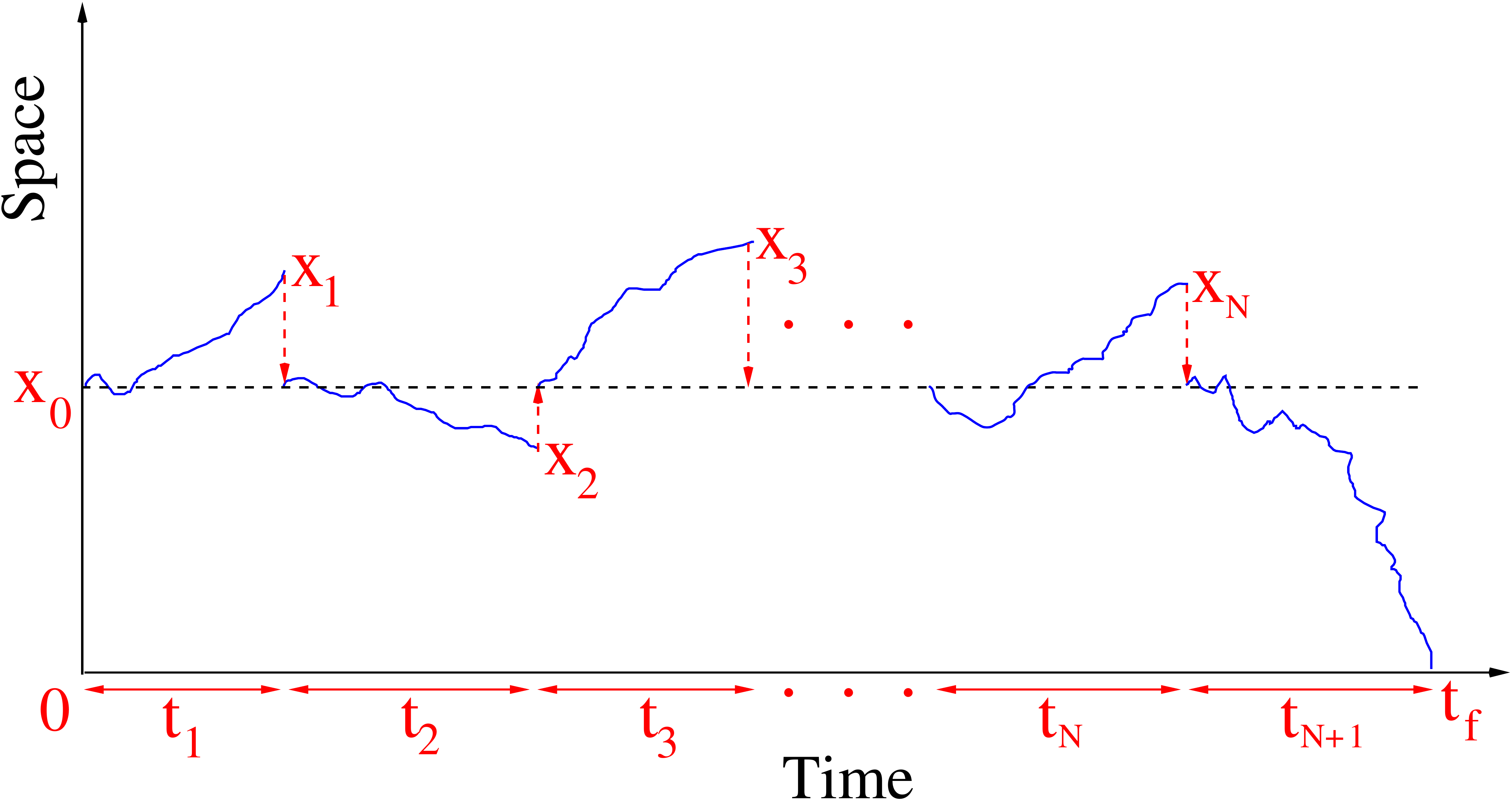}
    \caption{Example trajectory of diffusive process with resetting to $x=x_0$ and absorbing boundary at $x=0$. $t_1$ is the time of the first reset, $t_i$ is the time elapsed between $(i-1)^{\rm th}$ and $i^{\rm th}$ reset, and $t_f$ is the first passage time at which the particle is absorbed. $x_i$ is the position of the particle just before the $i^{\rm th}$ reset. Note that $t_{N+1}$ is the time elapsed between the $N^{\rm th}$ (final) reset and first passage time $t_f$. }
    \label{fig:schem}
\end{figure}
 where $\Theta$ is the Heaviside step function which constrains the cost to be non-negative. Equation \eqref{renew} can be understood by referring to Figure~\ref{fig:schem}.  The integration is over $t_1$, the time of the first reset,
  and  $x_1$, the position just before the first reset. The factor $ r {\rm e}^{-rt_1}{\rm d} t_1$ is the probability that the first reset occurs in the interval between $t_1$ and $ t_1 +{\rm d} t_1$ and $ G_0(x_1, t_1|x_0){\rm d} x_1 $ is the probability, in the absence of resetting,
for the process to have survived and have a position 
  between $x_1$ and  $x_1+ {\rm d} x_1$ at time $t_1$. 
  In principle, the cost of the first reset $c_1$ could be a general function of $x_1$, the position of 
  the particle just before the reset, and $x_0$, the resetting position. For clarity, we take the form where the cost  depends on the distance from the  resetting position, that is, $c_1=c(|x_1-x_0|)$, although much of the formalism below could be extended to the more general case (see  \cite{PKR20}). After the first reset, the process is renewed and  $F(N-1,t_f - t_1,C -c_1|x_0)$ is now the required joint probability  distribution function with the contribution of the first reset subtracted from the variables $N$, $t_f$ and $C$. Equation
  \eqref{renew} holds for $N>0$. For $N=0$ the corresponding equation is 
  \begin{equation}
    F(0,t_f,C|x_0) =  {\rm e}^{-rt_f}
                       F_0( t_f|x_0) \delta(C-0)\;, \label{renew1}
    \end{equation}
  where  $ {\rm e}^{-rt_f}$ is the probability of no resets up to time $t_f$, $F_0( t_f|x_0)$ is the probability density of completion at time $t_f$ in the absence of resetting and the total cost $C$ must be zero as there are no resets.

 We now take the Laplace transform of \eqref{renew} with Laplace variables $s$ and $p$ for $t_f$ and $C$ respectively. The resulting expression, derived in \ref{Appendix_LT_renewal}, is
 \begin{eqnarray}
   {\widetilde F}(N,s,p|x_0) &=& \int_0^\infty {\rm d} t_f\,  {\rm e}^{-st_f}\int_0^\infty {\rm d} C\, {\rm e}^{-pC}
                                 F(N,t_f,C|x_0)  \\
                             &=& r W(r+s,p|x_0)  {\widetilde F}(N-1,s,p|x_0)\;, \label{renewlt}
 \end{eqnarray}
 where $W(r+s,p|x_0)$ is defined as
\begin{align}
    W(r+s,p|x_0) = \int_0^\infty{\rm d}x\, {\rm e}^{-pc\left(\abs{x_0-x}\right)}\widetilde{G}_0(x,r+s|x_0)\;.
    \label{W_expression}
\end{align}
In \eqref{W_expression}, $\widetilde{G}_0(x,s|x_0)$ is the Laplace transform of $G_0(x,t|x_0)$ with respect to $t$
\begin{equation}
\widetilde{G}_0(x,s|x_0) = \int_0^\infty {\rm d}t\,
{\rm e}^{-st} G_0(x,t|x_0)\;.
\end{equation}
Finally we obtain from \eqref{renew1}
\begin{equation}
 {\widetilde F}(0,s,p|x_0) =  {\widetilde F}_0(r+s|x_0) 
\end{equation}
where $\widetilde{F}_0(s|x_0)$ is the Laplace transform of the first passage time distribution for the system without resetting.
Iterating \eqref{renewlt} yields our central result
 \begin{equation}
   {\widetilde F}(N,s,p|x_0) = 
                             \left[ r W(r+s,p|x_0)\right]^N  {\widetilde F}_0(r+s|x_0)\;. \label{Flt}
 \end{equation}
The importance of \eqref{Flt} is that it gives us the full joint statistics of the number of resets, the time to absorption and  the cost of resetting. For example, setting $p=s=0$
yields the distribution of the number of resets, $P(N| x_0)$  up to the first passage time, $t_f$:
\begin{equation}
   P(N|x_0) =\left[ r \widetilde{Q}_0(r|x_0)\right]^N  {\widetilde F}_0(r|x_0)\;, \label{PN}
 \end{equation}
where 
\begin{equation}
    W(r,0|x_0)=\widetilde Q_0(r|x_0) \label{Wp0}
\end{equation}
is the Laplace transform, with Laplace variable $r$, of the survival probability without resetting,  having started from $x_0$.
The distribution \eqref{PN} is evidently a geometric distribution and $r\widetilde{Q}_0(r|x_0)$ is simply the probability that the particle is not absorbed before the next reset. The geometric distribution \eqref{PN} is a simple  property of a renewal process and similar distributions appear in other contexts, such as  the probability of $N$ records in record statistics
\cite{MZ08}.

Similarly, setting $p=0$, which corresponds to integrating out the cost, and summing over $N$ recovers the Laplace transform, or equivalently moment generating function, of the first passage time distribution under resetting with rate $r$ \cite{EM1}
\begin{align}
\langle {\rm e}^{-s t_f} \rangle = \frac{{\widetilde F}_0(r+s|x_0)}
{1-r \widetilde{Q}_0(r+s|x_0)}\;. \label{Frlt}
 \end{align}
This has been used to study the mean first passage time and its minimization with respect to resetting rate \cite{EM1} and also the coefficient of variation \cite{Reuveni16,PR17} and various other properties.

We now use \eqref{Flt} to investigate the statistics of the total cost of resetting $C$. Setting $s=0$, which corresponds to integrating over all absorption times $t_f$, and summing over the number of resets $N$ gives the moment generating function of the cost distribution,
\begin{align}
\langle {\rm e}^{-pC}\rangle   =\frac{\widetilde{F}_0(r|x_0)}{1-rW(r,p|x_0)}\;.
\label{LT_cost_distribution}
\end{align}
The moments of the cost function are obtained as follows,
\begin{align}
    \begin{split}
        \langle C^n\rangle = (-1)^n\frac{\partial^n}{\partial p^n}\langle {\rm e}^{-pC}\rangle  \bigg|_{p \to 0^+}\;.
    \end{split}
\end{align}
The expressions derived so far are independent of the underlying process. The approach can also be generalized to higher dimensions as we will discuss in the conclusions. If we are provided with the first passage time density and the propagator for the system without resetting, we can, in principle derive the joint probability of number of resets, time of absorption and the cost. 

\subsection{The case of one-dimensional diffusion with resetting}
In the following, the underlying process we consider is one-dimensional diffusion with an absorbing boundary (the target) at $x=0$. We use the standard expressions for the Laplace transform of first-passage time density, Laplace transform of the propagator,
and Laplace transform of survival probability \cite{Redner,BMS13}:
\begin{align}
    \widetilde{F}_0(r|x_0) &= {\rm e}^{-\alpha_0 x_0},\label{F0}\\
    \widetilde{G}_0(x,r|x_0) &= \frac{1}{2\sqrt{Dr}}\left( {\rm e}^{-\alpha_0\abs{x-x_0}} - {\rm e}^{-\alpha_0\abs{x+x_0}} \right),\label{G0}\\
    \widetilde Q_0(r|x_0) &=\frac{1}{r}(1- {\rm e}^{-\alpha_0 x_0}),\label{Q0}
\end{align}      
where $\alpha_0 = \sqrt{r/D}$. The first two moments  of the total cost of resetting are then given by
\begin{align}
        \langle C\rangle &= -r{\rm e}^{\alpha_0 x_0}\frac{\partial}{\partial p}W(r,p|x_0)\bigg|_{p \to 0^+},\label{Cav}\\
        \langle C^2 \rangle  &=  \langle C \rangle ^2
        + r{\rm e}^{\alpha_0 x_0}\frac{\partial^2}{\partial p^2}W(r,p|x_0)\bigg|_{p \to 0^+}.
    \label{Cvar}
\end{align}
Using \eqref{Q0} and \eqref{F0} in \eqref{PN}, the distribution of the number of resets for one-dimensional diffusion with resetting becomes the geometric distribution
\begin{align}
    \begin{split}
        P(N|x_0) = {\rm e}^{-\gamma}(1-{\rm e}^{-\gamma})^N\;, \label{geometric_dist}
    \end{split}
\end{align}
where $\gamma$ is the dimensionless resetting rate
\begin{equation}
  \gamma =\sqrt{\frac{r}{D}}x_0\;.
  \label{gamdef}
\end{equation}
$\gamma$ is a key parameter that is used throughout this paper and can be interpreted as the ratio of two distances, namely the distance to the target, $x_0$ and the typical distance travelled between resets, $\sqrt{D/r}$.
\section{Linear resetting cost} \label{sec:linear}
\label{sec:lin}
\begin{figure}[h]
    \centering
    \includegraphics[width = 0.63\textwidth]{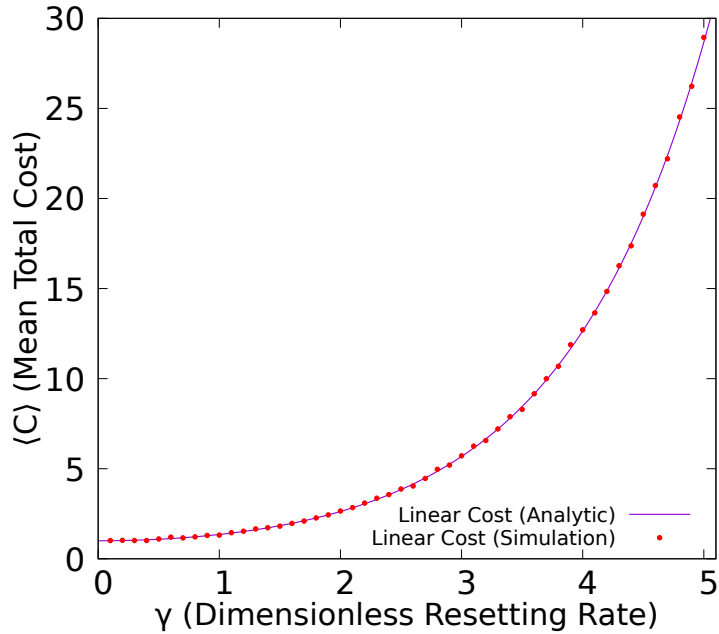}
    \caption{Simulation results of mean total cost obtained for a system with linear cost per reset (see \ref{Appendix_simulation} for details) compared with the analytical result obtained from the calculations. We have fixed $D=1,x_0=1$ and $V=1$.}
    \label{fig:linear_cost_simulation}
\end{figure}

As a first example, we consider a linear cost per reset, that is
\begin{equation}
  c_i = \frac{|x_i-x_0|}{V} \;. \label{Clin}
\end{equation}
Here $V$ has dimensions of velocity and is introduced to allow an interpretation of the cost as the time required to reset the process to its starting point. In this case, one finds from \eqref{W_expression} that
\begin{align}
    W(r+s,p|x_0) = \frac{1}{2\sqrt{D(r+s)}}\left( \frac{2}{\alpha+\hat{p}} + \frac{2\hat{p}}{\alpha^2 - \hat{p}^2} {\rm e}^{-2\alpha x_0} -\frac{2\alpha}{\alpha^2 - \hat{p}^2}{\rm e}^{-\left(\alpha + \hat{p}\right)x_0}\right)\;,
\end{align}
where $\hat p = p/V$ and $\alpha = \sqrt{(r+s)/D}$.
The mean of the total cost is then obtained from \eqref{Cav} as
\begin{align}
    \langle C\rangle _{\text{lin}} = \frac{x_0}{V}\left(\frac{2\sinh{\gamma}-\gamma}{\gamma}\right)\;.
    \label{linc}
\end{align}
The result \eqref{linc} has been obtained for a linear resetting cost, or home return time, in \cite{PKR20,BS20a,FPSRR20}. We have also verified the mean total cost using simulations---see Figure~\ref{fig:linear_cost_simulation}.

	%
   %

The mean total cost is minimized at  $\gamma^*$, which  is obtained by evaluating ${\rm d} \langle C \rangle/{\rm d} \gamma = 0$. For the linear case, it is clear that $\gamma^*_{\text{lin}} \to 0$, {\it i.e.} the limit of zero resetting rate. However, in this limit, the mean total cost is non-zero
\begin{equation}
        \langle C \rangle^*_{\text{lin}} = \frac{x_0}{V}
      \end{equation}
as opposed to a system with no resetting in which this cost would be zero. The finite mean total cost in the $r\to 0$ limit of the linear cost case has been noted in \cite{FPSRR20}. 

We now explain how this finite contribution to the total cost, in the limit of zero resetting rate, comes from a few rare trajectories of the process that do involve a reset.
The reset of these trajectories involves a large displacement of the particle, incurring a large cost. This interplay of a tiny, but significant resetting chance with the large displacement, which scales as the inverse of the small probability, contributes to the finite total mean total cost for resetting.

This argument can be made more rigorous by finding the probability of resets and the mean distance travelled between resets. The geometric distribution \eqref{geometric_dist}, in the limit of small rate of resetting {\it i.e.} $\gamma \to 0$ becomes
\begin{align}
    P(N|x_0) \simeq (1-\gamma)\gamma^N\;.
\end{align}
This distribution can be used to write down the leading contributions to the resetting events, which would be the case of no resets and a single reset to the leading order 
\begin{align}
    \begin{split}
        P(0|x_0) &= 1-\gamma + \mathcal{O}(\gamma^2)\;,\\
         P(1|x_0) &= \gamma + \mathcal{O}(\gamma^2)\;.
    \end{split}
    \label{resetting_order}
\end{align}

We now make an approximation  which ignores the absorbing boundary and consider the probability distribution of the position at the reset, 
\begin{align}
    P(x) &\simeq r\int_0^\infty dt \,    \frac{{\rm e}^{-rt}}{\sqrt{4\pi D t}} \exp\left( -\frac{(x-x_0)^2}{4Dt} \right) \label{dist_first_reset_pos}\\
         &= \frac{\alpha_0}{2} {\rm e}^ {- \alpha_0 |x-x_0|}\;.\label{dist_first_reset_pos_final}
\end{align}
 In \eqref{dist_first_reset_pos} we have used a Gaussian distribution for the position after time $t$,  and averaged over the distribution of times to the reset, employing a standard integral (see \cite{EMS20} Equation 2.17). In this way, we approximate the  distribution for the position just before the reset  with the exponential distribution \eqref{dist_first_reset_pos_final}. When the cost is averaged over this exponential we obtain
 \begin{equation}
    \langle c(|x-x_0|) \rangle \simeq \frac{1}{V}\int_{-\infty}^\infty {\rm d}x\,
    \frac{\alpha_0}{2} |x-x_0| {\rm e}^ {- \alpha_0 |x-x_0|}
     = \frac{1}{V}\,\frac{x_0}{\gamma}\;.
\end{equation}
The expected total cost for small $r$ is then obtained by multiplying the probability of the number of resets and the mean cost for the given number of resets
\begin{align}
\begin{split}
    \langle C\rangle_{\rm lin}  &\simeq  (1-\gamma)\cdot 0 + \gamma \cdot \frac{1}{V} \frac{x_0}{\gamma} + \mathcal{O}(\gamma) \\
    &=\frac{x_0}{V} + \mathcal{O}(\gamma) \;.
\end{split}
\end{align}

Thus as mentioned before, the non-zero contribution to the cost comes from the rare trajectories (with probability $\sim \gamma$) in which one reset occurs. That reset incurs a large cost $c \sim 1/\gamma$, resulting in a finite contribution to the mean total cost.

We can also consider the resetting rate which optimizes $\langle {\mathcal T} \rangle$, defined through \eqref{Tdef} as the mean total cost of resetting plus the  mean first passage time 
\begin{equation}
 \langle {\mathcal T}\rangle =\langle C \rangle_{\rm lin} + \langle t_f \rangle  \;,\label{Tlin} 
\end{equation}
where the mean first passage time is given by \cite{EMS20,EM2}
\begin{align}
    \langle t_f \rangle = \frac{x_0^2}{D} \left( \frac{{\rm e}^{\gamma}-1}{\gamma^2} \right)\;.
\label{mfpt}
\end{align}
This expression for $\langle t_f \rangle$  can  be obtained from \eqref{Frlt} by taking the derivative with respect to $s$ at $s=0$.
The first term in \eqref{Tlin} is an increasing function of $\gamma$ whereas the second term has a unique minimum at a non-zero value of $\gamma$.
Therefore the optimal resetting rate, which minimizes \eqref{Tlin} is {\em lower} than that in the absence of a cost.  
This optimal resetting rate is the solution to the transcendental equation
\begin{equation}
\frac{2\gamma^2\cosh(\gamma) - 2 \gamma \sinh(\gamma)} {\gamma {\rm e}^\gamma-2{\rm e}^\gamma+2} + \frac{x_0 V}{D} = 0\;.
\end{equation}

A similar expression for the combined minimization of mean total cost and mean first passage time has been obtained earlier \cite{PKR20}.

\section{Quadratic resetting cost} \label{sec:quad}

\begin{figure}[h]
    \centering
    \includegraphics[width = 0.63\textwidth]{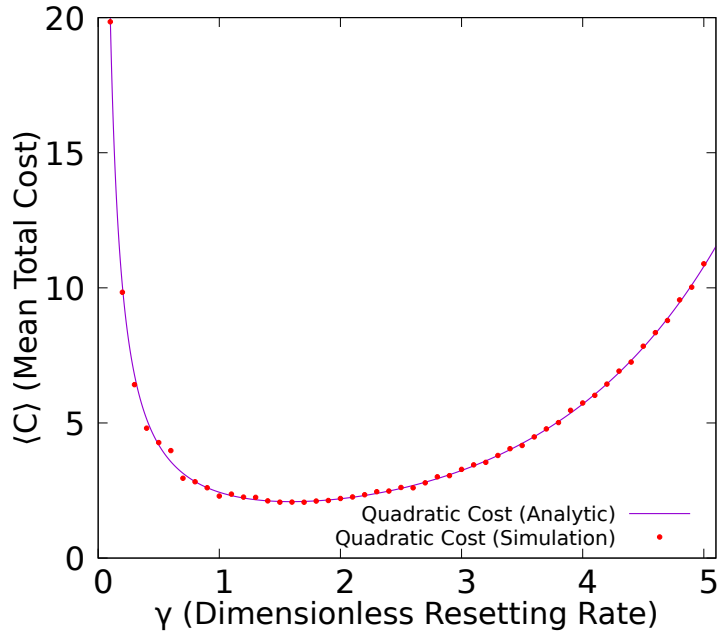}
    \caption{Simulation results of mean total cost obtained for a system with quadratic cost per reset (see \ref{Appendix_simulation} for details) compared with the analytical result obtained from the calculations. We have fixed $D=1,x_0=1$ and $V=1$.}
    \label{fig:quadratic_cost_simulation}
\end{figure}

We turn now to the case of  a quadratic cost per reset
\begin{align}
    c_i = \frac{\abs{x_i-x_0}^2}{V}\;. \label{Cquad}
\end{align}
A quadratic cost function has appeared in the context of entropy production of resetting in the stationary state of the system without an absorbing boundary \cite{MOK23}.
The exact expression for $W$ is provided in \ref{Appendix_W_expressions} \eqref{Wquadapp}. 
The mean total cost can be obtained for \eqref{Cquad}, using \eqref{Wquadapp} and \eqref{Cav}, as 
\begin{align}
   \langle C \rangle_{\text{quad}} = \frac{2x^2_0}{V} \left( \frac{{\rm e}^{\gamma}-1}{\gamma^2} \right)-\frac{x_0^2}{V}\;.
   \label{mean_quad_cost}
\end{align}
Expression \eqref{mean_quad_cost} is  in excellent agreement with simulation results, see Figure \ref{fig:quadratic_cost_simulation}.
Apart from a constant factor and an additive constant,
expression \eqref{mean_quad_cost} is of the form of the mean first passage time \eqref{mfpt}. Minimizing the mean quadratic cost thus results in the exact same transcendental equation as that of minimizing the mean first passage time \cite{EMS20,EM2}
\begin{align}
    \frac{\gamma}{2} = 1-{\rm e}^{-\gamma}
\end{align}
which can be solved to obtain $\gamma^* = 1.5936\ldots$.

This matching of optimal resetting rates is not a coincidence and can be explained using a diffusive scaling argument. The mean squared displacement between resets, ignoring the absorbing boundary, is given by the diffusive result
\begin{align}
    \langle (x_i-x_0)^2 \rangle = 2 D t_i \;.
\end{align}
Using this, we make the approximation
\begin{align}
    c(|x_i-x_0|) \simeq \frac{2D}{V}t_i\;. \label{C_quad_approx}    
\end{align}
Then given $N$ resets before absorption, the mean  total cost can be written as
\begin{align}
       \langle C \rangle_{\text{quad}} \simeq \frac{2D}{V}\left\langle \sum_{i=1}^N t_i \right\rangle
         \simeq \frac{2D}{V} \langle t_f \rangle\;, \label{quad_scaling_argument}
\end{align}
where $\langle t_f \rangle$ is given by \eqref{mfpt}. This approximation  recovers  \eqref{mean_quad_cost} up to an additive constant.
 One might wonder whether the equivalence between cost and $t_f$ extends beyond the mean. We have computed the variance of the cost, but a simple relation to the variance of $t_f$  does not hold and so the quadratic cost distribution is different in general from the distribution of $t_f$.

Further, upon inspecting the small resetting limit of (\ref{mean_quad_cost}), we see that the mean total cost diverges as
\begin{align}
    \langle C \rangle_{\text{quad}} \sim \frac{2}{\gamma}\left(\frac{x_0^2}{V}\right) \;.
\end{align}
This divergence can again be explained in terms of the number of resets and a scaling argument for the mean squared displacement. The number of resets in the limit $\gamma \to 0$ has a leading contribution of the order $\gamma$, while the mean squared displacement has a leading contribution of the order $1/\gamma^2$. The product of these, which gives the mean total quadratic cost, diverges as $1/\gamma$. Between the linear resetting cost and quadratic resetting cost, the behaviour of the mean total cost in the small resetting limit transitions from a finite value to an infinite value.

\section{General power resetting cost} \label{sec:power}
\begin{figure}[h]
    \centering
    \includegraphics[width = 0.63\textwidth]{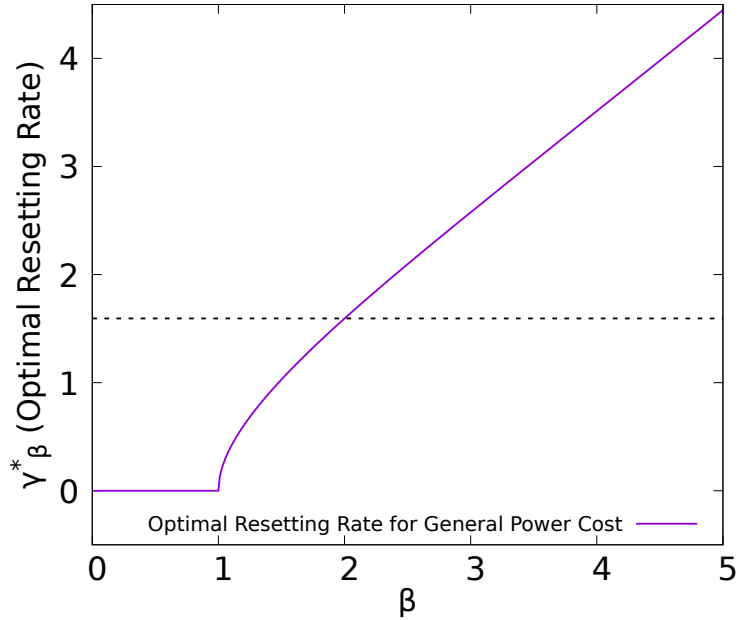}
    \caption{Optimal reset value for general $\beta$-power cost function ($\gamma^*_{\beta}$). We see a transition from zero resetting to a finite value of resetting at $\beta=1$. The optimal resetting rate in the absence of a cost, $\gamma_f = 1.5936\ldots$ is indicated by the dashed line}
    \label{beta_optimal_reset}
\end{figure}

To probe further the transition of the cost in the small resetting rate regime,  we now consider a generalization to the linear and quadratic cost by considering a general power cost, given by
\begin{align}
    c_i = \frac{\abs{x_i-x_0}^\beta}{V} \;. \label{beta_cost}
\end{align}
The case $\beta = 1/2$ corresponds to home returns with constant acceleration studied in \cite{BS20a}.
The mean total cost in the case  of general $\beta$ can be calculated as
\begin{align}
    \langle C \rangle_\beta = \frac{x_0^\beta}{V}\frac{\gamma^{-\beta}}{2} \bigg\lbrace &\left[2\sinh(\gamma)+{\rm e}^\gamma -(-1)^{-\beta-1}{\rm e}^{-\gamma}\right]\Gamma(\beta+1) + (-1)^{-\beta-1} {\rm e}^{-\gamma}\Gamma (\beta +1,-\gamma )\nonumber\\&-{\rm e}^{\gamma}\Gamma (\beta +1,\gamma ) \bigg\rbrace\;,
   \label{meanC_beta}
\end{align}
where $\Gamma(a,z)$ is the upper incomplete Gamma function given by
\begin{align}
    \Gamma(a,z) = \int_z^\infty dt\, t^{a-1}{\rm e}^{-t} \;.
\end{align}
The relevant quantities for the calculation of \eqref{meanC_beta} are provided in \ref{Appendix_W_expressions}. Recall that $\gamma$ in \eqref{meanC_beta} is the dimensionless resetting rate given by \eqref{gamdef}. For positive integers \eqref{meanC_beta} can be simplified to 
\begin{equation}
  \langle C \rangle_\beta =
    \begin{cases}
      \frac{x_0^\beta}{V}\frac{\gamma^{-\beta}}{2} \beta ! \left[ 2{\rm e}^{\gamma} - \sum_{k=0}^{\beta}\frac{1}{k!}\left( (-\gamma)^k + \gamma^k \right)\right] & \text{if $\beta$ is even}\;,\\ \\
      \frac{x_0^\beta}{V}\frac{\gamma^{-\beta}}{2} \beta ! \left[ 4\sinh(\gamma) + \sum_{k=0}^{\beta}\frac{1}{k!}\left( (-\gamma)^k - \gamma^k \right)\right] & \text{if $\beta$ is odd}\;.
    \end{cases}       
    \label{meanC_beta_integer}
\end{equation}
It is straightforward to check  that for values $\beta =1$ and $\beta = 2$, \eqref{meanC_beta_integer} recovers the linear and quadratic 
cost cases studied in sections~\ref{sec:lin} and \ref{sec:quad}. To study the small resetting rate, we use the following series expansion for $\Gamma(a,z)$
\begin{align}
    \Gamma(a,z) = \Gamma(a) - \sum_{k=0}^{\infty} \frac{(-1)^k z^{a+k}}{k!(a+k)} \qquad a \ne 0,-1,-2 \ldots
    \label{inc_gamma_expansion}
\end{align}
and obtain the cost in the limit $\gamma \to 0$ from \eqref{meanC_beta}, which is given by
\begin{align}
    \langle C \rangle_\beta \xrightarrow{\gamma \to 0}\frac{x_0^\beta}{V}\frac{\Gamma(1+\beta)}{\gamma^{\beta-1}} \;. \label{beta_cost_limit}
\end{align}
We can see from \eqref{beta_cost_limit} that there is a phase transition in the mean total cost in the $\gamma \to 0$ limit. The mean total cost undergoes a transition from 0 cost for $\beta<1$, to a finite value at $\beta=1$, to a divergence of the form $\gamma^{1-\beta}$ for $\beta>1$. The same behaviour can be observed in the optimal resetting rate as a function of $\beta$ (See Figure \ref{beta_optimal_reset}) which is obtained as the solution of the transcendental equation,
\begin{align}
        {\rm e}^{2\gamma } (\beta -\gamma ) \Gamma (\beta +1,\gamma )-\left(2 \beta  {\rm e}^{2 \gamma }- \gamma -2\gamma {\rm e}^{2 \gamma }  -\beta +(-1)^{-\beta }\left(\beta +\gamma \right)\right) \Gamma (\beta +1)\nonumber\\+(-1)^{-\beta }(\beta +\gamma ) \Gamma (\beta +1,-\gamma )=0 \;.
    \label{beta_optimal_rate}
\end{align}
The transition can again be explained using \eqref{resetting_order} which states that a reset has probability $\mathcal{O}(\gamma)$. The mean total cost has a finite value only when the vanishing resetting probability is compensated exactly by the cost given by (\ref{beta_cost}) which is of $\mathcal{O}(1/\gamma^\beta)$ which can be obtained using \eqref{dist_first_reset_pos_final}. This occurs only when $\beta=1$. For $\beta >1$ we obtain a divergent contribution.

We now turn to the optimization of the mean time to completion \eqref{Tdef}, which is the sum of the mean total cost and the mean first passage time.
We recall that the mean total cost \eqref{meanC_beta}
is minimized at the solution of \eqref{beta_optimal_rate} which we denote $\gamma_\beta^*$. The mean first passage time, on the other hand, has a unique minimum at $\gamma_f= 1.5936\ldots$.
Now, we have seen that for $\beta >2$, $\gamma_\beta^* > \gamma_f$.
Thus
${\rm d} \langle {\mathcal T}\rangle/{\rm d}\gamma\,<0$ at $\gamma = \gamma_f$
and
${\rm d} \langle {\mathcal T}\rangle/{\rm d}\gamma\,>0$ at $\gamma = \gamma_\beta^*$.
Therefore, the
mean time to completion (the sum of the mean total cost and mean first passage time) must have a minimum at a value of $\gamma$,  $\gamma_f< \gamma < \gamma_\beta^*$ when $\beta >2$. Thus we deduce that  a power law cost of resetting with power $\beta >2$ implies an {\em increased} optimal resetting rate over the case of no cost.

\section{Exponential resetting cost}
\label{sec-exp}

\begin{figure}[h]
    \centering
    \includegraphics[width = 0.63\textwidth]{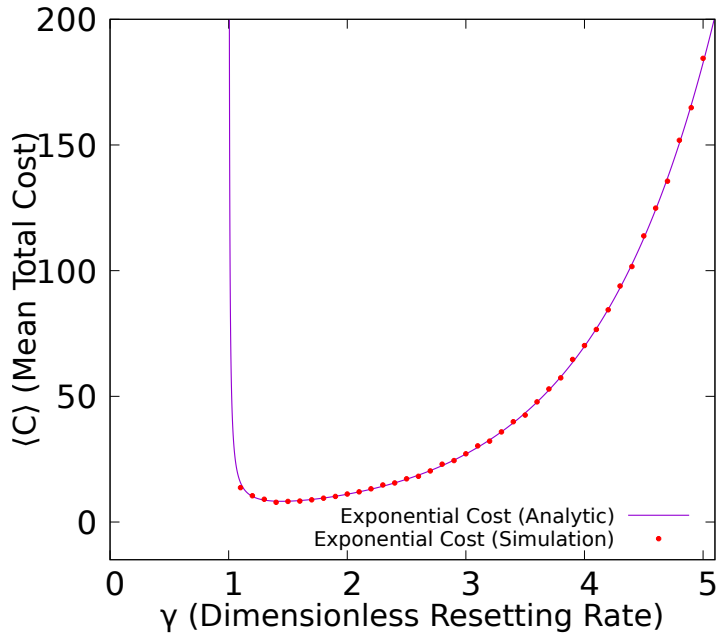}
    \caption{Simulation results of mean total cost obtained for a system with exponential cost per reset (see \ref{Appendix_simulation} for details) compared with the analytical result obtained from the calculations. We have fixed $\delta=1,D=1,x_0=1$ and $V=1$. The mean total cost diverges as discussed in the text for $\gamma\delta=1$.}
    \label{fig:exp_cost_simulation}
\end{figure}

\begin{figure}[h]
    \centering
    \includegraphics[width = 0.63\textwidth]{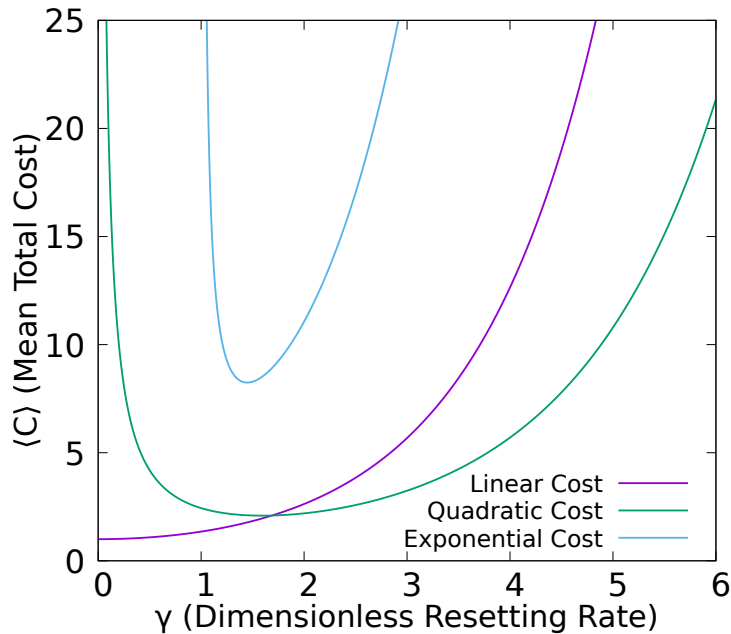}
    \caption{Plots of the various mean total costs. The mean total linear cost (Purple) has a non-zero finite limiting value in the small resetting limit while the mean total quadratic cost (Green) has a divergence in the same limit. The mean total exponential cost (Blue) has a divergence for a non-zero, finite value of the resetting rate.}
    \label{various_cost}
\end{figure}

\begin{figure}[h]
    \centering
    \includegraphics[width = 0.63\textwidth]{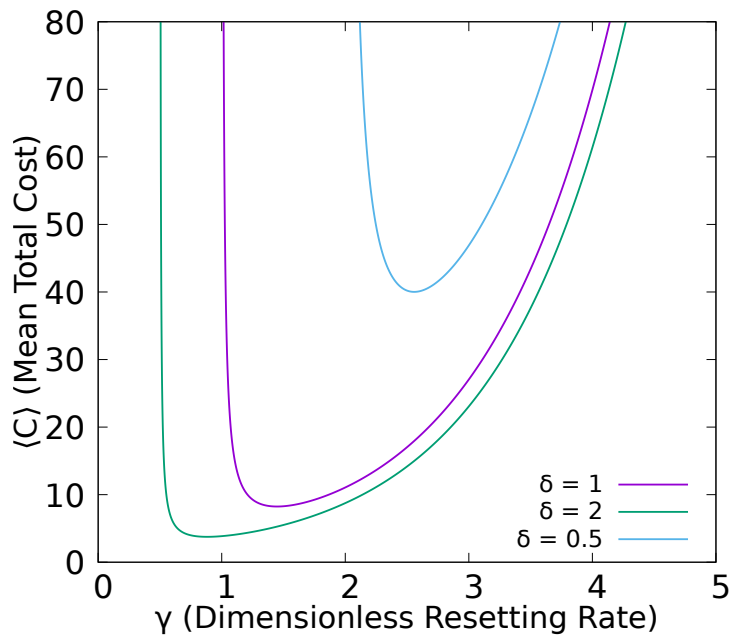}
    \caption{Plots of mean total exponential cost as a function of $\gamma$ for different choices of $\delta$. We see the divergence at $\gamma = 1/\delta$ as discussed in the text.}
    \label{various_exp_cost}
\end{figure}

Finally, we consider an exponentially increasing cost per reset
\begin{align}
    c_i = \frac{1}{V}{\exp\left( \frac{\abs{x_i-x_0}}{\kappa} \right)}\;, \label{exp_cost}
\end{align}
where $\kappa$ is the scaling length for the cost. The motivation here is that, due to the exponential function growing faster than any power, it models situations where the particle  being transported a long distance are heavily penalized. For example, this could be because of physical constraints limiting the distance of transport. By defining a dimensionless length scale
\begin{equation}
    \delta = \kappa/x_0
    \end{equation}
 and using the dimensionless resetting rate $\gamma$ \eqref{gamdef}, the mean total cost is obtained from \eqref{Cav} as
\begin{align}
    \langle C \rangle_{\text{exp}} = \frac{\gamma \delta}{V} \left(\frac{2\sinh(\gamma)+\gamma \delta \left({\rm e}^{\gamma} - {\rm e}^{\frac{1}{\delta}} \right)}{\gamma^2\delta^2-1}\right) \qquad \text{when} \quad \gamma>\frac{1}{\delta}\;.
    \label{Cexp}
\end{align}
For an exact expression of the function $W$ for this case, see \ref{Appendix_W_expressions}.
Expression \eqref{Cexp} is compared with simulation results in Figure \ref{fig:exp_cost_simulation}.
The agreement is excellent for $\gamma \delta >1$ and there are some fluctuations due to finite sampling error near $\gamma \delta =1$.
Compared to all the previous cases, the exponential cost is  interesting as the mean total cost diverges for a non-zero  resetting rate as shown in Figure~\ref{various_cost}. Figure~\ref{various_exp_cost} illustrates that the mean total cost diverges at $\gamma \delta =1$.

\begin{figure}[h]
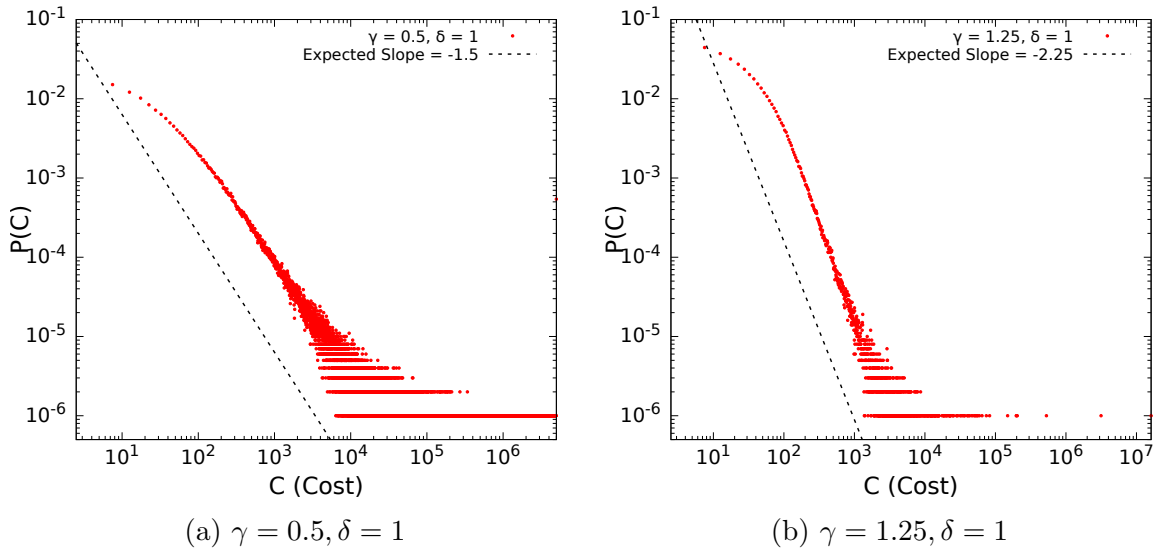

    \centering
	\begin{subfigure}{.49\textwidth}
		\centering
		\includegraphics[width=\linewidth]{power_law_distribution_g_0_5.pdf}
		\caption{$\gamma=0.5,\delta=1$}  
		\end{subfigure}
		\begin{subfigure}{.49\textwidth}
		\centering
	    \includegraphics[width=\linewidth]{power_law_distribution_g_1_25.pdf} 
		\caption{$\gamma=1.25,\delta=1$} 
	\end{subfigure}
    \caption{Log-log plots of the probability density function of total exponential cost obtained from simulations. The expected slope of the data computed analytically is plotted alongside for comparison.}
    \label{power_law_distribution_cost}
    \end{figure}

In \ref{Appendix_power_law} we derive the exact large $C$ asymptotic behaviour  for the distribution of the total cost and find
\begin{align}
     P(C) \sim  \frac{1}{V^{\gamma\delta}}\gamma\delta \sinh(\gamma) C^{-(\gamma\delta+1)}\;.
     \label{Cpl}
 \end{align}
The power-law tail obtained in (\ref{Cpl}) indicates that the mean total cost is divergent for $\gamma \delta <1$ and variance is divergent for $\gamma \delta<2$. We probed this power-law behaviour using simulations (See Figure \ref{power_law_distribution_cost}) which are consistent with the prediction \eqref{Cpl}.

The  power-law distribution and divergence of the mean total cost for $\gamma \delta <1$ can be explained in a simple way by 
using the approximate  probability distribution of the position at the reset \eqref{dist_first_reset_pos_final} (obtained by ignoring the absorbing boundary). When the exponential cost per reset is averaged over this exponential position distribution, we get a divergence when $ \alpha_0 < 1/\kappa$ ({\it i.e.} $\gamma <1/\delta$). This essentially corresponds to a competition between the exponentially decaying tail of the steady-state distribution and the exponentially growing contribution of the cost. The transition from finite to infinite cost occurs at the point where the contributions from both are exactly the same {\it i.e.} $\gamma = 1/\delta$.
We can also consider the distribution of the cost per reset, $f(c)$, defined as
\begin{align}
f(c) = P(x)\abs{\frac{dx}{dc}}\;.
\end{align}
Then using approximation \eqref{dist_first_reset_pos_final} and
\begin{align}
    c = \frac{1}{V}{\exp\left( \frac{\abs{x-x_0}}{\kappa} \right)}\;,
\end{align}
one finds
\begin{align}
f(c) \simeq \frac{\gamma \delta}{2V^{\gamma \delta}} c^{-(\gamma\delta+1)}\;.  \label{power_law}
\end{align}
Thus the combination of an exponential tail of the distribution  of resetting  distances and an exponential resetting cost gives rise to a power-law decay of the cost per reset distribution with a continuously varying exponent
$\gamma\delta +1$.
The mechanism leading to this power-law decay appears in many contexts such as trap models of glassy dynamics \cite{Bouchaud92}, where the time to exit a trap scales as $\exp\left(E/(k_B T)\right)$, and the trap energy $E$ has an exponential  distribution $P(E)\sim \exp\left(-  E/\bar{E}\right)$. Consequently, the waiting time in a trap has a power-law distribution. Another similar mechanism can be found in 
\cite{DDHM05}.

The distribution of the mean total cost \eqref{Cpl} is consistent with $C$ being a sum of random variables each distributed according to \eqref{power_law}. To understand this, we note that
$C$, defined in \eqref{Cdef}, is a sum of $N$ random variables $c_i$, where $N$ is the number of resets.
In the large  $C$  limit  the probability of the sum taking value $C$ is dominated by one of the random variables taking value close to  $C$ and  the others
taking typical values. This phenomenon is known as condensation 
and the single dominant random variable is known as the condensate \cite{EMZ06}. Then the probability of the large $C$ event is  given by $P(C) \simeq Nf(c= C)$ where the factor $N$ comes from the $N$ possibilities for the condensate.
In the present case $N$, the number of resets,
is itself a random variable with geometric distribution \eqref{geometric_dist}. The mean of $N$ is then given by ${\rm e}^\gamma$. 
We find ${\rm e}^\gamma f(c=C)$ where $f(c)$ is given by \eqref{power_law} indeed recovers  the exact asymptotic \eqref{Cpl} of $P(C)$ when $\gamma$ is large so that
$\sinh \gamma \simeq {\rm e}^\gamma/2$.

\begin{figure}[h]
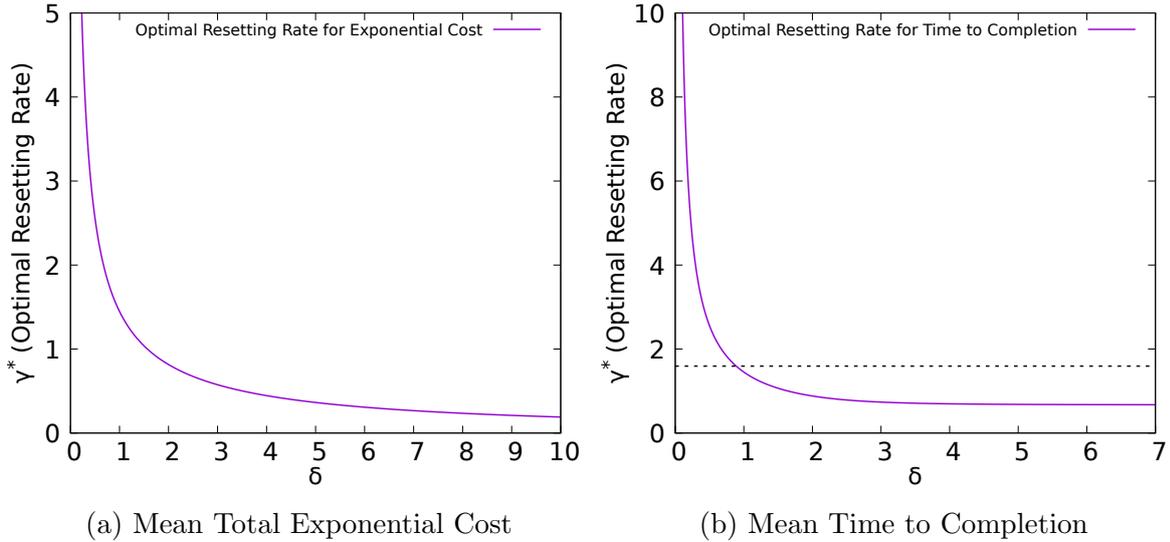

    \begin{subfigure}{.49\textwidth}
    \centering
    \includegraphics[width=\textwidth]{Optimal_Resetting_Rate_for_Exponential_Cost.pdf}
    \caption{Mean Total Exponential Cost}
    \label{exp_optimal_reset}
    \end{subfigure}
	\begin{subfigure}{.49\textwidth}
		\centering
	    \includegraphics[width=\linewidth]{Optimal_Resetting_Rate_for_Total_Exponential_Cost.pdf} 
		\caption{Mean Time to Completion} 
		\label{exp_optimal_completion}
	\end{subfigure}

    \caption{Optimal reset value in the case of an exponential cost per reset \eqref{exp_cost} as a function of $\delta$ for (a) the mean total  cost ($\langle C\rangle_{\text{exp}}$) and (b) the mean time to completion ($\mathcal{T})$ . For (b) the case of mean time to completion, we have chosen $D=x_0=V=1$, and we see that the optimal resetting rate can be greater than or lesser than the optimal rate for the first passage problem (Black dashed line) depending on the value of $\delta$.} 
\end{figure}

Further, the optimal resetting rate for the minima of the mean total exponential cost can be obtained for a given value of $\delta$ as the solution to the transcendental equation
\begin{align}
    \gamma(\gamma^2\delta^2-1)\cosh(\gamma) -(\gamma^2\delta^2+1)\sinh(\gamma)+\gamma\delta\left( \frac{\gamma^3\delta^2}{2} - \frac{\gamma}{2}-1\right){\rm e}^\gamma + \gamma \delta {\rm e}^{\frac{1}{\delta}} = 0 \;.
\end{align}
We plot the solution against
$\delta$ in Figure \ref{exp_optimal_reset}. We also show in Figure \ref{exp_optimal_completion} that by varying $\delta$, the resetting rate which minimizes the mean time to completion may be increased or decreased with respect to the rate which minimizes the mean first passage time, $\gamma_f$.

\section{Conclusions}

In this work, we have studied the mean total cost of resetting for one-dimensional diffusion using an additive cost function with various forms for the cost of an individual reset.
We  have derived the Laplace transform of the joint probability distribution of the number of resets, first passage time and total cost function, from which all moments can be computed.

We have considered several different forms for the cost per reset: linear, power law and exponential. 
When we optimize the time to completion \eqref{Tdef} (sum of cost plus first passage time) we find that compared to optimising only the first passage time, the optimal resetting rate is reduced in the linear cost case, unchanged in the quadratic cost case and  increased
for power-law costs with exponent greater than two.
For exponentially increasing cost per reset, the optimal resetting rate may be
decreased or increased with respect to the no cost case. What is perhaps surprising is that the introduction of a cost can increase the optimal resetting rate, but the explanation is that this reduces the chance of large displacement resets.
Thus the subtle interplay between the minimization of the absorption time for a task and minimization of the cost of resets must be borne in mind when designing resetting strategies for more general problems.

Other unexpected phenomena have emerged, namely that in the limit of zero resetting rate, the mean total cost remains finite for a linear cost function.  Moreover, for super-linear cost functions the mean total cost is divergent in this limit. This counter-intuitive behaviour can be understood as the effect of resetting events with vanishing probability making divergent contributions to the total cost.

In addition, we have seen that for an exponential cost function, the mean  total cost diverges at  a non-zero value of resetting. This results from an exponentially decreasing tail of the reset displacement competing with an exponentially increasing cost per reset.
This interplay results in a power-law distribution for the cost per reset. The presence of a power-law distribution suggests that rare events dominate the cost in this regime and it would be interesting to explore this effect in more detail.

In this study, we have limited ourselves to one spatial dimension. However, it is relevant to consider higher dimensions as well. In particular, two dimensions may be relevant to experimental realizations of resetting. The relevant expressions \eqref{renew} and \eqref{renew1} generalize in higher dimensions to
\begin{eqnarray}
    F(N,t_f,C|\bm{x_0}) &=& \int_0^{t_f} {\rm d} t_1 r {\rm e}^{-rt_1} \int_0^\infty {\rm d}^d \bm{x_1}
                       G_0(\bm{x_1}, t_1|x_0) \\
    &&\times F(N-1,t_f - t_1,C -c_1|\bm{x_0}) \Theta(C-c_1)\;,  \nonumber
  \end{eqnarray}
with
\begin{equation}
    F(0,t_f,C|\bm{x_0}) =  {\rm e}^{-rt_f}
                       F_0( t_f|\bm{x_0}) \delta(C-0)\;.
    \end{equation}
However for a diffusive process with an absorbing target in dimensions $d \geq 2$, one has to modify the target to have a finite radius $a$. Relevant details can be found in \cite{EM14}.

Another simple extension would be to calculate the long-time cost to maintain a nonequilibrium steady state, $P_{ss}(x)$, that is induced by resetting. This involves an average cost per unit time $\langle\mathcal{C}\rangle_\text{ss}$, which in the case of a Poissonian resetting process would simply be
\begin{align}
    \langle \mathcal{C}\rangle_{\text{ss}} = r\int_{-\infty}^{\infty}{\rm d}x\;c(\abs{x-x_0})P_{ss}(x)\;.
\end{align}

This equation expresses $\langle \mathcal{C}\rangle_{\text{ss}} $ as the average cost per reset in the steady state times the resetting rate.  As the stationary state for diffusion under Poissonian resetting is given by an exponential
$P_{ss}(x) = (\alpha_0/2) {\rm e}^{-\alpha_0 |x-x_0|}$,
we obtain
\begin{align}
    \langle \mathcal{C}\rangle_{\text{ss}} = r \alpha_0 {\cal L}_{\alpha_0}\lbrace c\rbrace\;,
\end{align}
where ${\cal L}_{\alpha_0}$ denotes Laplace transform with Laplace variable $\alpha_0$.
As in \sref{sec-exp}, there is a divergence in the case of an exponential cost per reset when $r$ is sufficiently small.

It would also be of  interest to extend  the considerations of the cost  of resetting to other propagators than that of diffusion with an absorbing boundary. For example, anomalous diffusion processes where distance travelled and time scale as
\begin{align}
    \langle (x-x_0)^2 \rangle &\sim t^{\mu}\;,
\end{align}
where $\mu$ may take values different from unity, have been studied under resetting in
\cite{MPCM19,KGN19,MM19}
and it should be possible to extend calculations to the cost of resetting.

\section*{Acknowledgements}
We thank Somrita Ray for the helpful discussions. JCS acknowledges the award of EDCS from the University of Edinburgh. MRE thanks LPTMS for the award of a Visiting Professorship. For the purpose of open access, the authors have applied a Creative Commons Attribution (CC BY) licence to any Author Accepted Manuscript version arising from this submission. 

\appendix
\section{Laplace transform of renewal equation} \label{Appendix_LT_renewal}
We start from the definition of the renewal equation \eqref{renew} and perform a double Laplace transform with respect to $t_f$ and $C$. The variable conjugate to $t_f$ is $s$ and the variable conjugate to $C$ is $p$. We then get
\begin{eqnarray}
\fl        \widetilde{F}(N,s,p|x_0) &=& \int_0^\infty {\rm d}t_f\, {\rm e}^{-st_f} \int_0^\infty {\rm d}C\,{\rm e}^{-pC}\int_0^{t_f} {\rm d} t_1\, r {\rm e}^{-rt_1} \int_0^\infty {\rm d} x_1\,
                       G_0(x_1, t_1|x_0) \nonumber\\
    &&\times F(N-1,t_f - t_1,C -c_1|x_0) \Theta(C-c_1)\\
    &=&r\int_0^\infty {\rm d}C\,{\rm e}^{-pC} \int_0^\infty {\rm d} x_1\, \int_0^\infty {\rm d}t_f\, {\rm e}^{-st_f} \int_0^{t_f} {\rm d} t_1  {\rm e}^{-rt_1} 
                       G_0(x_1, t_1|x_0)\nonumber\\
    &&\times F(N-1,t_f - t_1,C -c_1|x_0) \Theta(C-c_1)\;,
    \label{LT_renewal}
\end{eqnarray}
where
\begin{equation}
    c_1 = c(\abs{x_0-x_1})\;.
\end{equation}
 The last two integrals in \eqref{LT_renewal} constitute the Laplace transform of a convolution integral, which can be simplified using the convolution theorem for Laplace transforms as
\begin{align}
    \int_0^\infty {\rm d}t_f\, {\rm e}^{-st_f} \int_0^{t_f} {\rm d} t_1\,  {\rm e}^{-rt_1} 
                       G_0(x_1, t_1|x_0) F(N-1,t_f - t_1,C -c_1|x_0) \nonumber \\
                       = \left[\int_0^\infty{\rm d}t_1\, {\rm e}^{-(r+s)t_1}G_0(x_1, t_1|x_0)\right] \left[ \int_0^\infty{\rm d}t_2\, {\rm e}^{-st_2} F(N-1,t_2,C -c_1|x_0)\right]\nonumber \\
                       =  \widetilde{G}_0(x_1, r+s|x_0)\widehat{F}(N-1,s,C -c_1|x_0)\;. \label{single_LT}
\end{align}
$\widehat{F}(N-1,s,C -c_1|x_0)$ in \eqref{single_LT} represents the single Laplace transform with respect to time.
Collecting the remaining terms and making a change in variable $C^\prime = C - c_1$ gives
\begin{align}
     \widetilde{F}(N,s,p|x_0) &= r \int_0^\infty {\rm d}x_1\,\int_{-c_1}^\infty {\rm d}C^\prime\, {\rm e}^{-p(C^\prime + c_1)} \widetilde{G}_0(x_1, r+s|x_0)\nonumber\\ &\quad\times \widehat{F}(N-1,s,C^\prime|x_0) \Theta(C^\prime)\\
    &= r \left[ \int_0^\infty {\rm d}x_1\, {\rm e}^{-pc_1}\widetilde{G}_0(x_1, r+s|x_0) \right] \nonumber \\ &\quad \times \left[ \int_{-c_1}^\infty dC^\prime\, {\rm e}^{-pC^\prime}\widehat{F}(N-1,s,C^\prime|x_0) \Theta(C^\prime)\right]\\ 
    &=  r W(r+s,p|x_0) \left[ \int_0^\infty dC^\prime\, {\rm e}^{-pC^\prime}\widehat{F}(N-1,s,C^\prime|x_0) \right]\\
    & = r W(r+s,p|x_0) \widetilde{F}(N-1,s,p|x_0)\;,
 \end{align}
where we have used \eqref{W_expression} to rewrite the integral over $x_1$ as $W(r+s,p|x_0)$.

\section{Exact expressions for $W(r,p|x_0)$} \label{Appendix_W_expressions}
The expression for $W(r,p|x_0)$ is evaluated by substituting \eqref{G0} in \eqref{W_expression},
\begin{align}
    W(r,p|x_0) = \int_0^\infty{\rm d}x\, {\rm e}^{-pc\left(\abs{x-x_0}\right)}
    \frac{1}{2 D \alpha_0}\left( {\rm e}^{-\alpha_0\abs{x-x_0}} - {\rm e}^{-\alpha_0\abs{x+x_0}} \right)\;,
        \label{W}
\end{align}
where $\alpha_0 = \sqrt{r/D}$, and carrying out the integral wherever possible. 
In order to obtain $W(r+s,p|x_0)$ one simply replaces $\alpha_0$ with $\alpha = \sqrt{(r+s)/D}$ in \eqref{W} and subsequent expressions.

In cases, where the calculation of $W(r,p|x_0)$ is difficult, we can still obtain the derivatives of $W(r,p|x_0)$ with respect to $p$ easily by switching the order of integration and differentiation in (\ref{W_expression}). 
\subsection{Linear Cost}
In the case of linear cost per reset \eqref{Clin}
\begin{equation}
  c(|x-x_0|) = \frac{|x-x_0|}{V} \;, \label{Clinapp}
\end{equation}
we obtain
\begin{align}
    W(r,p|x_0) = \frac{1}{2 D \alpha_0}\left( \frac{2}{\alpha_0+\hat{p}} + \frac{2\hat{p}}{\alpha_0^2 - \hat{p}^2} {\rm e}^{-2\alpha_0 x_0} -\frac{2\alpha_0}{\alpha_0^2 - \hat{p}^2}{\rm e}^{-\left(\alpha_0 + \hat{p}\right)x_0}\right)\;,
    \label{Wlin}
\end{align}
where  $\hat{p} = p/V$,
and 
\begin{align}
    \frac{\partial}{\partial p}W(r,p|x_0)\bigg|_{p \to 0^+} &= -\frac{1}{V}\frac{1}{2D \alpha_0} \left[ \int_0^\infty dx \abs{x-x_0} {\rm e}^{-\alpha_0\abs{x-x_0}} - \int_0^\infty dx \abs{x-x_0} {\rm e}^{-\alpha_0\abs{x+x_0}}\right] \nonumber \\
    &= \frac{1}{V} \frac{1}{D} \frac{{\rm e}^{-\alpha_0 x_0}}{\alpha_0^3} \left(\alpha_0 x_0-2\sinh(\alpha_0 x_0)\right)\nonumber \\
    &= \frac{x_0^3}{V D} \frac{{\rm e}^{-\gamma}}{\gamma^3} \left(\gamma-2\sinh(\gamma)\right)\;.
\end{align}

\subsection{Quadratic Cost}
In the case of quadratic cost per reset \eqref{Cquad}
\begin{align}
    c( |x-x_0|) = \frac{\abs{x-x_0}^2}{V}\;, \label{Cquadapp}
\end{align}
we obtain
\begin{align}
    W(r,p|x_0) = \frac{1}{4 D\alpha_0}\sqrt{\frac{\pi}{\hat{p}}}{\rm e}^{\frac{\alpha_0^2}{4\hat{p}}}\Bigg[\erf\left({\sqrt{\hat{p}}x_0+\frac{\alpha_0}{2\sqrt{\hat{p}}}}\right)-2\erf\left( \frac{\alpha_0}{2\sqrt{\hat{p}}}\right)\nonumber \\
    -{\rm e}^{-2\alpha_0 x_0} \erf\left({\sqrt{\hat{p}}x_0-\frac{\alpha_0}{2\sqrt{\hat{p}}}}\right)+2{\rm e}^{-\alpha_0 x_0}\sinh\left(\alpha_0 x_0\right) \Bigg]\;,
\label{Wquadapp}
\end{align}
\begin{align}
    \frac{\partial}{\partial p}W(r,p|x_0)\bigg|_{p \to 0^+} &= -\frac{1}{V}\frac{1}{2 D \alpha_0} \left[ \int_0^\infty dx (x-x_0)^2 {\rm e}^{-\alpha_0\abs{x-x_0}} - \int_0^\infty dx (x-x_0)^2 {\rm e}^{-\alpha_0\abs{x+x_0}}\right] \nonumber\\
    &= \frac{1}{V} \frac{1}{D} \frac{{\rm e}^{-\alpha_0 x_0}}{\alpha_0^4} \left( -2{\rm e}^{-\alpha_0 x_0} +\alpha_0^2x_0^2 + 2\right)\nonumber\\
    &= \frac{x_0^4}{V D} \frac{{\rm e}^{-\gamma}}{\gamma^4} \left( -2{\rm e}^{-\gamma} +\gamma^2 + 2\right)\;,
\end{align}
where $\hat{p} = p/V$ and $\alpha_0 = \sqrt{r/D}$. We have used the usual definition of the error function given by
\begin{align}
    \erf(x) = \frac{2}{\sqrt{\pi}} \int_0^x {\rm e}^{-t^2}\;dt\;.
\end{align}

\subsection{General power cost}
In the case of a power-law cost per reset \eqref{beta_cost},
\begin{align}
    c(|x-x_0|) = \frac{\abs{x-x_0}^\beta}{V} \;, \label{beta_cost_app}
\end{align}
a closed-form computation of $W(r,p|x_0)$ is not possible for general $\beta$. However, the derivative with respect to $p$ in the limit $p \to 0^+$ can always be obtained. We then get the expression
\begin{align}
        \frac{\partial}{\partial p}W(r,p|x_0)\bigg|_{p \to 0^+} &= -\frac{1}{V}\frac{1}{2D \alpha_0} \left[ \int_0^\infty dx \abs{x-x_0}^\beta {\rm e}^{-\alpha_0\abs{x-x_0}} - \int_0^\infty dx \abs{x-x_0}^\beta {\rm e}^{-\alpha_0\abs{x+x_0}}\right]\nonumber\\
    &= \frac{1}{V}\frac{\alpha _0^{-\beta -2} {\rm e}^{-\alpha _0 x_0}}{2D} \Bigg\lbrace \Big[-2\sinh{(\alpha _0 x_0)} \Gamma (\beta +1) -{\rm e}^{\alpha _0 x_0} + \left(-1\right)^{-\beta -1} {\rm e}^{-\alpha _0 x_0}\Big]\nonumber\\
   &\quad\Gamma (\beta +1)+\left(-1\right)^{-\beta -1}{\rm e}^{-\alpha _0 x_0}\Gamma(\beta +1,-\alpha_0 x_0) - {\rm e}^{\alpha _0 x_0}\Gamma(\beta +1,\alpha_0 x_0)\Bigg\rbrace\nonumber\\
    &= \frac{x_0^{\beta+2}}{VD}\frac{ {\rm e}^{-\gamma}}{2\gamma^{\beta+2}} \Bigg\lbrace \Big[-2\sinh{(\gamma)} \Gamma (\beta +1) -{\rm e}^{\gamma} + \left(-1\right)^{-\beta -1} {\rm e}^{-\gamma}\Big]\Gamma (\beta +1)\nonumber\\
   &\quad-\left(-1\right)^{-\beta -1}{\rm e}^{-\gamma}\Gamma(\beta +1,-\gamma) + {\rm e}^{\gamma}\Gamma(\beta +1,\gamma)\Bigg\rbrace\;.
\end{align}
\subsection{Exponential cost}
In the case of an exponential cost per reset \eqref{exp_cost},
\begin{align}
    c(|x-x_0|) = \frac{1}{V}{\exp\left( \frac{\abs{x-x_0}}{\kappa} \right)}\;, \label{exp_costapp}
\end{align}
we obtain
\begin{align}
    \begin{split}
        W(r+s,p|x_0) = \frac{x_0^2}{D}\frac{\delta}{2 \gamma} \Bigg\lbrace& \hat{p}^{\gamma\delta} \left[ 2\Gamma(-\gamma\delta,\hat{p})-\Gamma(-\gamma\delta,\hat{p}{\rm e}^{\frac{1}{\delta}})-{\rm e}^{-2\gamma}\Gamma(-\gamma\delta,\hat{p})\right]\\&-\hat{p}^{-\gamma\delta}{\rm e}^{-2\gamma}\left[ \Gamma(\gamma\delta,\hat{p})-\Gamma(\gamma\delta,\hat{p}{\rm e}^{\frac{1}{\delta}}) \right]\Bigg\rbrace\;,
    \end{split}
    \label{W_exp}
\end{align}
\begin{align}
    \begin{split}
    \frac{\partial}{\partial p}W(r,p|x_0)\bigg|_{p \to 0^+} &= -\frac{1}{V}\frac{1}{2D\alpha_0} \left[ \int_0^\infty dx\; {\rm e}^{\frac{\abs{x-x_0}}{\kappa}} {\rm e}^{-\alpha_0\abs{x-x_0}} - \int_0^\infty dx\; {\rm e}^{\frac{\abs{x-x_0}}{\kappa}} {\rm e}^{-\alpha_0\abs{x+x_0}}\right]\\
    &= \frac{x_0^2}{VD} \frac{\delta}{\gamma}{\rm e}^{-\gamma} \left(\frac{-2\sinh(\gamma)-\gamma\delta \left({\rm e}^{\gamma} - {\rm e}^{\frac{1}{\delta}}\right)}{\gamma^2\delta^2-1}\right) \qquad \text{when} \quad \gamma>\frac{1}{\delta}\;,
    \end{split}
\end{align}
where $\hat{p}=p/V$ and $\delta = \kappa/x_0$.

\section{Simulation algorithm} \label{Appendix_simulation}
To circumvent the issue of time-expensive computation of the trajectories,  we use an algorithm that employs inverse transform sampling. This technique requires the cumulative distribution function of the reset position, which can be computed from the propagator of a diffusive system with an absorbing boundary. The schematic algorithm for the simulation of cost is given as follows:

\begin{itemize}
    \item Choose the time $\tau$ where the next reset happens from the exponential distribution with mean $1/r$.
    \item Terminate with probability
    \begin{align}
        P_A(\tau) = 1- \int_0^\infty dx\,P(x,t_A>\tau)\;,
    \end{align}
    where $P(x,t_A>\tau)$ is the propagator given by
    \begin{align}
        P(x,t_A>\tau) = \frac{1}{\sqrt{4 \pi D t}} \left[\exp\left( -\frac{(x-x_0)^2}{4Dt} \right) - \exp\left( -\frac{(x+x_0)^2}{4Dt} \right) \right]\;.
    \end{align}
    \item Sample the position $x$ where the next reset happens from the distribution
    \begin{align}
        R(x;\tau) = \frac{P(x,t_A>\tau)}{1-P_A(\tau)}\;,
    \end{align}
    where $P_A(\tau)$ is the probability that the particle has been absorbed at time $\tau$. Sampling from this distribution is accomplished by using the inversion method for which we need the cumulative distribution
    \begin{align}
        P(x>X,t_A>\tau) = \int_X^\infty{\rm d}x\, P(x,t_A>\tau) = \frac{1}{2} \left[ \erfc{\left(\frac{X-x_0}{\sqrt{4D\tau}} \right)} - \erfc{\left(\frac{X+x_0}{\sqrt{4D\tau}} \right)} \right],
    \end{align}
    and the absorption probability
    \begin{align}
        P_A(\tau) = 1- P(x>0,t_A>\tau) = \erfc{\left( \frac{x_0}{\sqrt{4D\tau}} \right)}\;.
    \end{align}
    \end{itemize}

 To summarize, we implement the above scheme with the following algorithm:
    \begin{itemize}
        \item Choose the time $\tau$ until the next reset from the exponential distribution with mean $1/r$.
        \item Choose a uniform random variable $u \in [0,1)$.
        \item Terminate if
        \begin{align}
            u \geq \erf{\left( \frac{x_0}{\sqrt{4D \tau}} \right)}\;.
        \end{align}
        \item Solve the equation
        \begin{align}
            \frac{1}{2} \left[ \erfc{\left(\frac{X-x_0}{\sqrt{4D\tau}} \right)} - \erfc{\left(\frac{X+x_0}{\sqrt{4D\tau}} \right)} \right] = \erf{\left( \frac{x_0}{\sqrt{4D \tau}}\right) - u }
        \end{align}
        for X which is the next reset position. This is solved using a numerical root-finding algorithm.
        \item Find the cost, for the function of choice for the given reset.
        \item Repeat till the process terminates.
    \end{itemize}
We performed the above steps for 10,000 runs, calculated the total cost for each trajectory till absorption and averaged the total cost over the total number of runs to generate the plots for the mean total cost of resetting (See Figures \ref{fig:linear_cost_simulation},\ref{fig:quadratic_cost_simulation} and \ref{fig:exp_cost_simulation}).

\section{Power law tail of exponential cost distribution} \label{Appendix_power_law}
Here we consider an exponential distribution for the cost per reset and  determine the large $C$ behaviour of the distribution of the total cost.
 To obtain the large cost behaviour, we Laplace invert (\ref{LT_cost_distribution}) which is the moment generating function for the cost or the Laplace transform of the cost distribution function for small arguments $p$. The branch cut at the origin due to the incomplete gamma function from (\ref{W_exp}) implies a power-law decay rather than an exponential decay for the large cost behaviour of the distribution of the cost. We will consider here the case of non-integer $\gamma \delta$. For integer $\gamma \delta$ we expect logarithmic corrections.
 
 We perform small $\hat p$ expansion  \eqref{W_exp} using (\ref{inc_gamma_expansion}), retain leading non-integer terms ${\hat p}^{\gamma \delta}$  and ignore integer order terms $O(p)$,$O(p^2)$ etc to obtain
\begin{align}
    rW(r,p|x_0) \simeq  (1-{\rm e}^{-\gamma}) + \frac{\gamma\delta}{2}\left( 1 - {\rm e}^{-2\gamma} \right)\Gamma(-\gamma\delta) \hat{p}^{\gamma\delta} \;,
\end{align}
 where $\hat{p} = p/V$. Expanding \eqref{LT_cost_distribution} we obtain
\begin{align}
\langle {\rm e}^{-pC}\rangle   &=\frac{\widetilde{F}_0(r|x_0)}{1-rW(r,p|x_0)}
\label{LT_cost_distribution_app}\\
    &\simeq 1 + \gamma \delta \sinh(\gamma)\Gamma(-\gamma\delta) \hat{p}^{\gamma\delta} \;,
\end{align}
where we have used $\widetilde{F}_0(r|x_0) = {\rm e}^{-\gamma}$, and again we have ignored integer order terms.

Formally inverting using the Laplace transform
 \begin{align}
     \mathcal{L}_p\lbrace C^q \rbrace = \frac{\Gamma(q+1)}{p^{q+1}}\;,
 \end{align}
 we get the power law tail behaviour as $C \to \infty$ for the total exponential cost function
 \begin{align}
     P(C) \sim \frac{1}{V^{\gamma\delta}}\gamma\delta \sinh(\gamma) C^{-(\gamma\delta+1)}\;.
 \end{align}

\end{document}